\documentclass[prb,reprint,amsmath,amssymb,superscriptaddress,floatfix]{revtex4-2}

\usepackage{graphicx}\usepackage{bm}\usepackage{booktabs}\usepackage{dcolumn}

\newcolumntype{d}[1]{D{.}{.}{#1}}

\usepackage{amsmath,amssymb,graphicx,xcolor}
\usepackage{graphicx}
\usepackage{bm}

\begin{document}

\title{Mass-asymmetry-controlled exciton dressing and dissociation in a quantum lattice model}

\author{Michael O. Atambo}
\email{michael.atambo@tukenya.ac.ke}
\affiliation{Department of Physics, Earth and Environmental Science, Technical University of Kenya, Nairobi, Kenya}
\date{\today}

\begin{abstract}
In a polar material, a neutral exciton couples to phonons through the sum of the electron and hole deformation potentials. Because the total source vanishes by charge neutrality, the elastic exciton-phonon vertex is regularized by electron-hole interference. Here we determine the non-perturbative fate of this interference by exact diagonalization of a Holstein-exciton model. By parameterizing the mass asymmetry to decouple it from the small-polaron atomic limit, we map a regime map comprising an internal dressing crossover and a dissociation boundary. We prove analytically and verify numerically that for equal masses, the symmetric exciton ground state is protected from phonon dressing by an exact exchange selection rule, provided the phonon source is odd under electron-hole exchange (the neutral case). As mass asymmetry increases, this selection rule is broken and the exciton acquires a strong local polaronic cloud. We show that the dissociation boundary, conversely, is set by a global energy balance between Coulomb binding and polaronic stabilization, and is nearly independent of the internal dressing. Paradoxically, the very symmetry that protects the exciton from dressing denies it polaronic stabilization, driving it toward dissociation at strong coupling. We discuss these results in the context of lattice exciton-polaron models and their implications for sharp versus broad excitonic lines in mass-symmetric versus mass-asymmetric polar semiconductors.
\end{abstract}

\maketitle

\section{Introduction}

The optical and transport properties of polar semiconductors are governed by the interaction of charge carriers and bound electron-hole pairs with lattice vibrations~\cite{landau1933,frohlich1954,holstein1959,toyozawa1958,rashba1982,emin2013,alexandrov1995}. An electron or hole coupled to phonons dresses into a polaron, and when the coupling exceeds the kinetic energy scale it may self-trap~\cite{song1993}. For a Coulomb-bound electron-hole pair the situation is subtler: a neutral exciton couples to phonons through the \emph{sum} of the electron and hole sources. Whether the bound excitation inherits the polaronic clouds of its constituents-becoming an exciton-polaron-or remains a bare, mobile excitation with a sharp zero-phonon line (ZPL) is the central question of this work.

Lattice models of exciton-polarons have a rich history. Hohenadler, Littlewood, and Fehske performed pioneering unbiased quantum Monte Carlo studies of the 1D lattice exciton-polaron, demonstrating that the relative sign and magnitude of the electron and hole couplings dictate the localization and binding of the pair~\cite{hohenadler2007}. Subsequent diagrammatic Monte Carlo work by Burovski \emph{et al.} provided numerically exact benchmarks for exciton-polaron formation with Coulomb attraction~\cite{burovski2008}. More recently, modern theoretical frameworks have connected these model Hamiltonians to first-principles calculations of excitonic polarons in real materials~\cite{dai2024,ghosh2023}, and ultrafast spectroscopy has revealed the dynamics of exciton-polaron formation in 2D perovskites~\cite{naturemat2024}. 

Despite this extensive literature, a fundamental question remains regarding the interplay between \emph{internal dressing} (how much the bound exciton interacts with phonons) and \emph{thermodynamic stability} (whether the bound exciton remains lower in energy than two separated polarons). Conventional wisdom often attributes the dichotomy between sharp excitonic lines in mass-symmetric halide perovskites~\cite{miyata2017,johnston2016} and broad, heavily dressed lines in mass-asymmetric oxides like ZnO and TiO$_2$~\cite{klingshirn2007,song1993} purely to the magnitude of the coupling constants. Here we show that the mass ratio of the constituents is an equally decisive control parameter, because it governs the quantum interference and exchange symmetry between the electron and hole.

In this work, we use finite-size exact diagonalization to isolate the effect of constituent mass asymmetry on exciton dressing. We introduce a parameterization that cleanly separates the symmetry-breaking effect of mass asymmetry from the trivial small-polaron localization effect. We derive an exact exchange selection rule that protects the mass-symmetric exciton from dressing, and we verify this theorem with explicit control calculations that break the source parity. Finally, we demonstrate that while mass asymmetry controls the internal dressing crossover, the dissociation boundary is governed by a largely mass-independent global energy balance. This leads to a striking conclusion: the symmetry protection that prevents dressing simultaneously denies the exciton polaronic stabilization, making it thermodynamically vulnerable to dissociation at strong coupling.

\section{Model and Symmetry Structure}\label{sec:model}

We study one electron and one hole on an $N$-site chain coupled to local Einstein phonons,
\begin{widetext}
\begin{equation}
H = -t_e\!\!\sum_{\langle ij\rangle}\! c_i^\dagger c_j - t_h\!\!\sum_{\langle ij\rangle}\! d_i^\dagger d_j - V\!\!\sum_i n_i^e n_i^h + \omega_0\!\!\sum_i b_i^\dagger b_i + \sum_i \left(g_e n_i^e - g_h n_i^h\right)\!(b_i^\dagger + b_i).
\end{equation}
\end{widetext}
To cleanly isolate the effect of mass asymmetry from the small-polaron atomic limit, we parameterize the hoppings as $t_e = t_0 e^\delta$ and $t_h = t_0 e^{-\delta}$. The geometric mean $\sqrt{t_e t_h} = t_0$ is held constant, fixing the ``reduced mass'' scale. The parameter $\delta$ controls the mass ratio $m_h/m_e = e^{2\delta}$ without driving either carrier into the strict $t \to 0$ localized limit. 

\begin{figure}[t]
\centering
\includegraphics[width=\columnwidth]{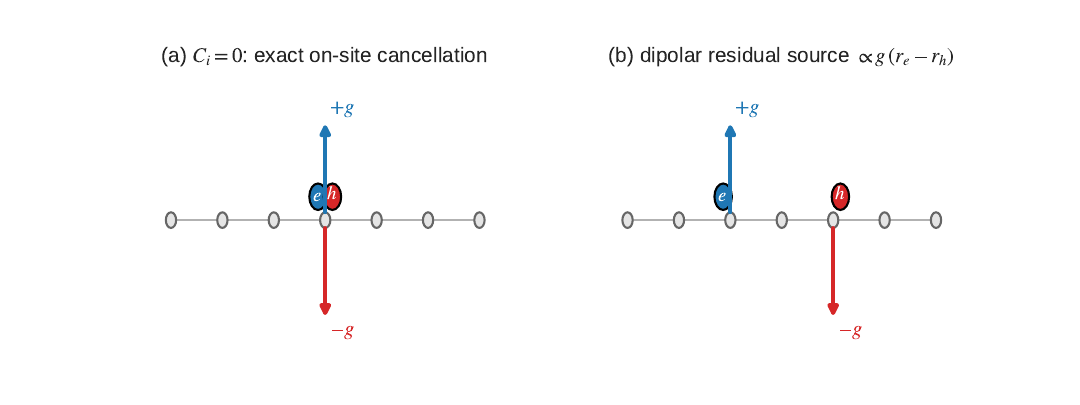} 
\caption{Phonon source and exchange symmetry. (a) For a coincident electron-hole pair, the neutral source $C_i = g(\delta_{i,r_e} - \delta_{i,r_h})$ vanishes site by site and is strictly odd under the electron-hole exchange operator $\mathcal{P}_{eh}$. (b) For separated carriers, a dipolar residual source remains. The odd parity of the neutral source, combined with the symmetric ground state at equal masses, enforces an exact exchange selection rule that suppresses phonon dressing.}
\label{fig:schematic}
\end{figure}

For a neutral exciton, we set $g_e = g_h \equiv g$. A configuration $(r_e,r_h)$ creates the local phonon source
\begin{equation}
C_i(r_e,r_h) = g\,(\delta_{i,r_e}-\delta_{i,r_h}).
\end{equation}
This source satisfies charge neutrality ($\sum_i C_i = 0$), but more importantly, it is strictly \emph{odd} under the electron-hole exchange operator $\mathcal{P}_{eh}$, which swaps $c_i \leftrightarrow d_i$. The structure of the source and its parity under exchange are illustrated
in Fig.~\ref{fig:schematic}.

\subsection{The Exchange Selection Rule}
At perfect mass symmetry ($\delta=0$, $t_e=t_h$), the unperturbed Hamiltonian $H_0 = H_{\rm kin} + H_{\rm ph} + H_{\rm Coulomb}$ is invariant under exchange: $[\mathcal{P}_{eh}, H_0] = 0$. Its non-degenerate ground state $|\Psi_0\rangle$ must therefore be symmetric: $\mathcal{P}_{eh}|\Psi_0\rangle = +|\Psi_0\rangle$. 

Because the interaction $H_{\rm int} = \sum_i C_i (b_i^\dagger + b_i)$ is odd under $\mathcal{P}_{eh}$, the first-order energy shift vanishes exactly:
\begin{equation}
\langle \Psi_0 | H_{\rm int} | \Psi_0 \rangle = - \langle \Psi_0 | H_{\rm int} | \Psi_0 \rangle = 0.
\end{equation}
Furthermore, coupling to any one-phonon state requires exciting the electronic relative coordinate into an antisymmetric state, which costs kinetic and Coulomb energy. This exchange selection rule rigorously suppresses the phonon dressing of the symmetric exciton. As we demonstrate below, this protection strictly requires the odd (neutral) parity of the source; an even source breaks the selection rule and leads to heavy dressing even at $\delta=0$.

\emph{Relation to concurrent works.}%
The selection rule established here concerns the \emph{diagonal}, first-order
coupling of the exciton ground state: because the phonon source is odd under
electron-hole exchange $\mathcal{P}_{eh}$, the expectation value
$\langle \Psi_0 | H_{\mathrm{int}} | \Psi_0 \rangle$ vanishes. The lattice activation law-the
statement that the lattice intraband coefficient equals $\tfrac{3}{2}\eta$-was
first derived in Ref.~\cite{B}.

The lattice coefficient $\tfrac{3}{2}\eta$ quoted in Ref.~\cite{B}
is the sum of two distinct pieces: a center-of-mass phase-twist piece $\eta$,
which bookkeeps COM recoil in the lattice COM gauge, and a quadrupole piece
$\tfrac{1}{2}\eta$, which coincides with the exact Galilean-invariant
continuum coefficient of Refs.~\cite{C}. The $a\to 0$ test of
Ref.~\cite{B}-intercept$/\Delta = 0.98$-$1.00$ across the Wannier-Frenkel
crossover-shows that the twist is a recoil partition rather than a genuine
finite-bandwidth correction. Either way, the diagonal selection rule proved
here is independent of this partition: it relies only on the odd parity of the
source under $\mathcal{P}_{eh}$, not on how the transition coefficient is
split between twist and quadrupole. Throughout we retain the sign convention
$\eta=(m_h-m_e)/(m_h+m_e)=(t_e-t_h)/(t_e+t_h)$ shared with
Refs.~\cite{B,C}, so that $\eta>0$ denotes the heavier hole.

\section{Method}\label{sec:method}

We diagonalize $H$ exactly in the sector $\{r_e,r_h,\{n_i\}\}$ with a total-phonon truncation $\sum_i n_i\le n_{\rm tot}$; for $N=8$, $n_{\rm tot}=6$ this gives $\dim\mathcal{H}=192{,}192$. Observables include the binding energy relative to independently converged free polarons, $E_{\rm bind}=E_e^{\rm pol}+E_h^{\rm pol}-E_X$; the zero-phonon weight $Z_0=\sum_{e,h}|\Psi(e,h;\{0\})|^2$, a ground-state measure of zero-phonon character; and the total phonon number $\langle N_{\rm ph}\rangle$. Convergence and the two-tier reference protocol are documented in
Appendix~\ref{app:conv} and Fig.~\ref{fig:convergence}.

\section{Results}\label{sec:results}

\subsection{Mass-asymmetry-controlled dressing}
\begin{figure}[t]
\centering
\includegraphics[width=\columnwidth]{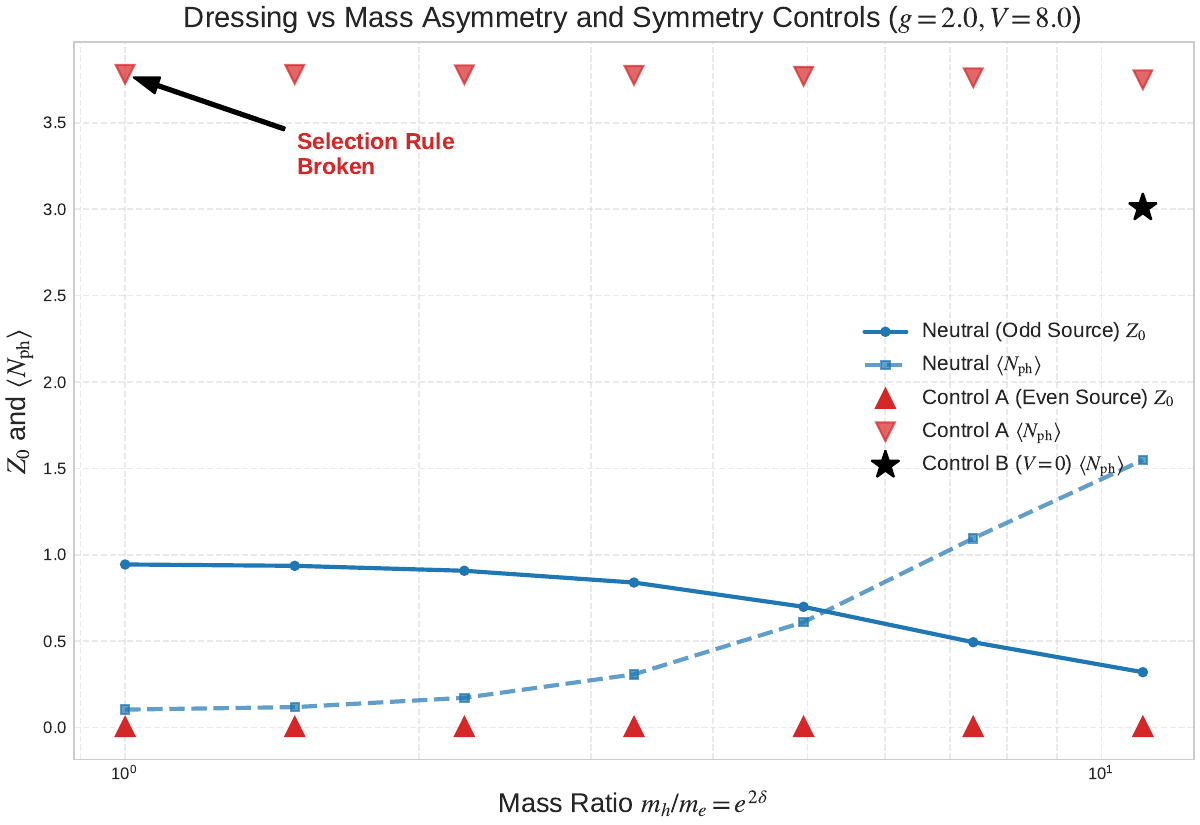} \caption{Mass-asymmetry-controlled dressing and symmetry controls at $g=2.0$, $V=8.0$. \textbf{Main curves:} Evolution of the zero-phonon character $Z_0$ and total phonon number $\langle N_{\rm ph}\rangle$ as the mass ratio $m_h/m_e = e^{2\delta}$ increases. The $\delta$-parameterization holds the geometric mean of hoppings constant, isolating the symmetry effect from the small-polaron atomic limit. The symmetric exciton ($\delta=0$) is protected by the exchange selection rule. \textbf{Control A (Even Source):} Reversing the hole coupling sign ($g_h = -g$) renders the source even under $\mathcal{P}_{eh}$, explicitly breaking the selection rule and causing heavy local polaronic dressing ($\langle N_{\rm ph}\rangle \approx 3.8$) even at perfect mass symmetry, consistent with the relative-sign physics identified in Ref.~\cite{hohenadler2007}. \textbf{Control B ($V=0$):} Removing the Coulomb bound state at high asymmetry leads to heavy dressing, confirming the suppression is a property of the correlated pair.}
\label{fig:delta_scan}
\end{figure}

Figure~\ref{fig:delta_scan} shows the evolution of the exciton as the mass ratio $e^{2\delta}$ is increased at fixed coupling $g=2.0$ and binding $V=8.0$. For the neutral (odd) source, the symmetric system ($\delta=0$) is highly protected: $Z_0=0.94$ and $\langle N_{\rm ph}\rangle=0.10$. As the mass asymmetry increases to a ratio of 11:1 ($\delta=1.2$), the selection rule is broken by the asymmetric kinetic energy, and the exciton acquires a strong local polaronic cloud ($Z_0=0.32$, $\langle N_{\rm ph}\rangle=1.55$). This smooth crossover demonstrates that mass asymmetry directly controls the internal dressing of the bound state.

\subsection{Control Calculations: Symmetry and Correlation}
To rigorously establish the origin of the protection at $\delta=0$, we perform two control calculations. 

\emph{Control A (Broken Source Parity):} We reverse the sign of the hole coupling ($g_h = -g$), rendering the phonon source $C_i = g(n_i^e + n_i^h)$ \emph{even} under $\mathcal{P}_{eh}$. The exchange selection rule is explicitly broken. As shown in Fig.~\ref{fig:delta_scan}, the symmetric exciton ($\delta=0$) immediately acquires a massive phonon cloud ($\langle N_{\rm ph}\rangle \approx 3.8$, $Z_0 \approx 0.005$). This proves that the protection observed for the neutral exciton is not merely a consequence of charge neutrality or mass symmetry, but specifically relies on the odd parity of the neutral source combined with the symmetric ground state.

\begin{figure}[t]
\centering
\includegraphics[width=\columnwidth]{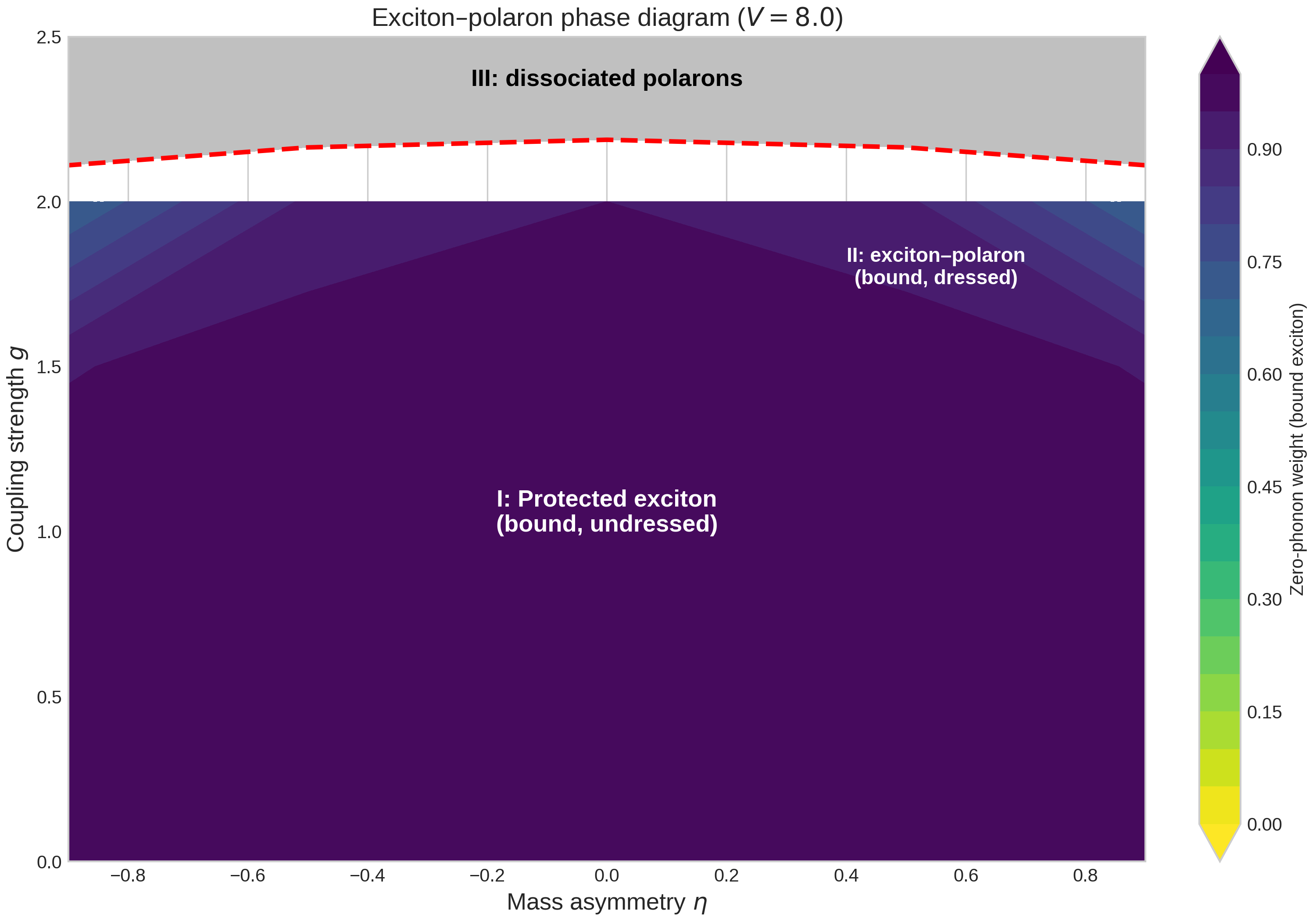} 
\caption{Exciton-polaron regime map at $V=8$: zero-phonon character $Z_0$ in the bound region (color), the dissociated region (gray), and the $E_{\rm bind}=0$ boundary (red dashed). The map highlights the separation of energy scales: mass asymmetry $\eta$ controls the internal dressing crossover (Regime I to II), while the global energetic competition between Coulomb binding and polaronic stabilization sets the nearly $\eta$-independent dissociation boundary (Regime III).}
\label{fig:regime_map}
\end{figure}

\emph{Control B (Uncorrelated Carriers):} We turn off the Coulomb attraction ($V=0$) at high asymmetry ($\delta=1.2$). Without the bound state, the system dresses heavily ($\langle N_{\rm ph}\rangle \approx 3.0$). This confirms that the partial suppression of dressing observed at finite $\delta$ for the bound exciton is a genuine property of the correlated pair resisting dissociation, rather than a feature of independent carriers. The corresponding two-parameter overview is shown in Fig.~\ref{fig:regime_map}:
the dressing crossover widens toward large mass ratios, while the
$E_{\rm bind}=0$ boundary remains essentially vertical, i.e. controlled by
$(g,V)$ rather than by the internal structure of the bound state.

\subsection{Dissociation and the Energy Balance}
While mass asymmetry controls the internal dressing crossover, the thermodynamic stability of the exciton is governed by a different mechanism. A free carrier gains polaronic stabilization $E_{\rm pol}\approx -g^2/\omega_0$, so two separated polarons lie at $\approx-2g^2/\omega_0$. Equating this to the bare exciton energy yields the dissociation estimate $g_c \approx \sqrt{\omega_0|E_X^{\rm bare}|/2}$.

\begin{figure}[t]
\centering
\includegraphics[width=\columnwidth]{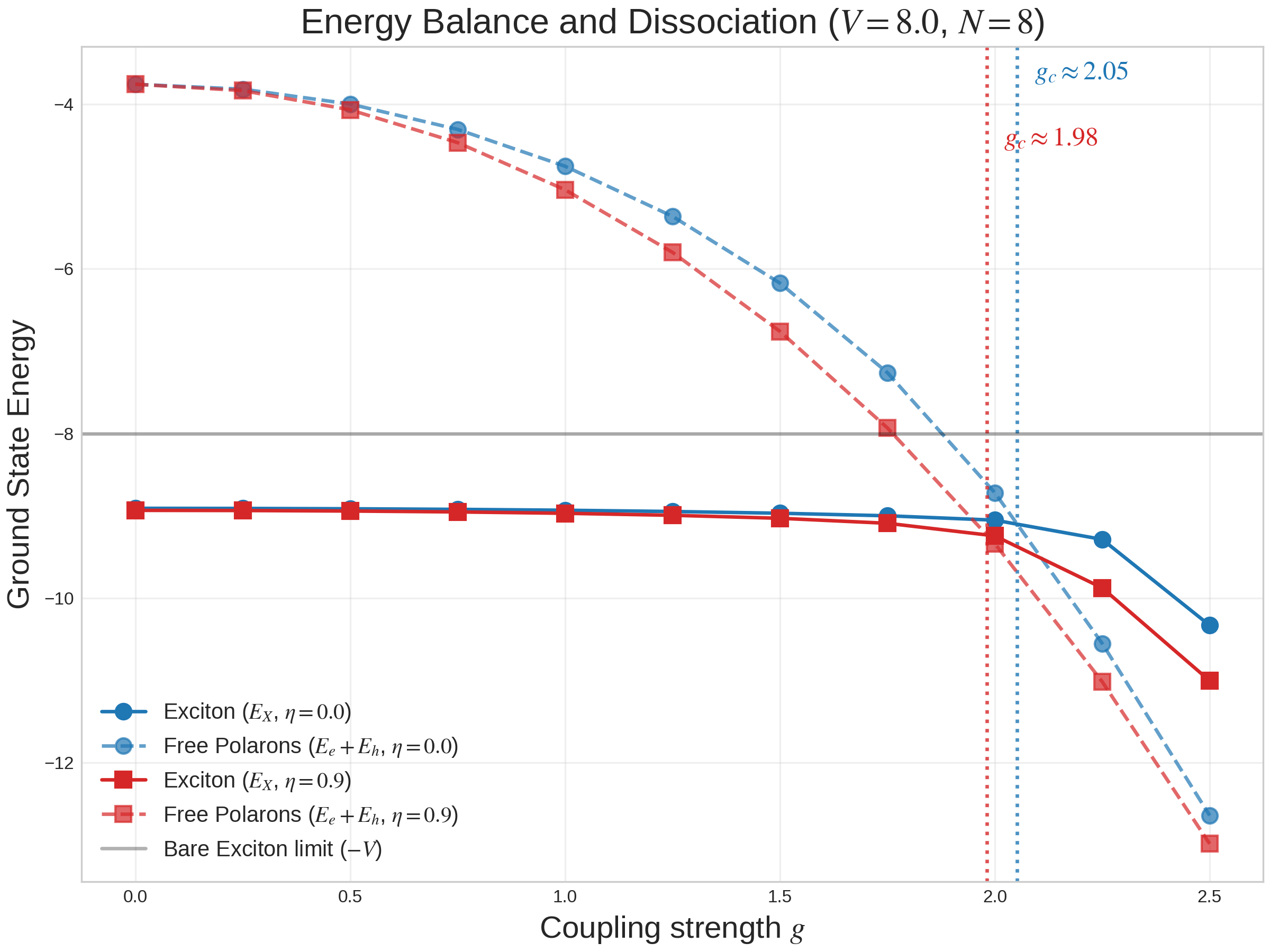} 
\caption{Energy-balance construction for the dissociation boundary at $V=8$. The exciton ground-state energy (solid) is compared to the sum of the independently converged free-polaron energies (dashed). The protected exciton remains remarkably flat, harvesting almost no polaronic stabilization, leading to a finite-size level crossing at $g_c \approx 2.05$ that is consistent with a dissociation transition in the thermodynamic limit. The crossing location is largely independent of the internal dressing state.}
\label{fig:balance}
\end{figure}

Figure~\ref{fig:balance} compares the exciton ground-state energy with the sum of the converged free-polaron energies. For the protected exciton ($\delta=0$), the energy remains remarkably flat as $g$ increases, harvesting almost no polaronic stabilization. The two-polaron branch descends rapidly, and the branches cross at $g_c \approx 2.05$. This finite-size level crossing is consistent with a dissociation transition in the thermodynamic limit. Beyond the crossing, the ground state switches to the dissociated branch. 

Crucially, the dissociation boundary is nearly independent of the mass asymmetry $\delta$ and the internal dressing. Increasing $V$ from 8 to 12 shifts the crossing to $g_c \approx 2.5$, exactly tracking the energy-balance prediction. 

\subsection{The Paradox of Protection}
These results reveal a striking paradox. The exchange symmetry that protects the mass-symmetric exciton from phonon dressing simultaneously denies it the polaronic stabilization energy that lowers the energy of the free carriers. Consequently, as the coupling $g$ increases, the dressed free polarons inevitably undercut the protected exciton. Protection suppresses dressing, but within this model, it makes dissociation energetically inevitable at sufficiently strong coupling.

\section{Discussion and Outlook}

The regime map established in this work clarifies the relationship between internal exciton dressing and global thermodynamic stability in quantum lattice models. Our findings connect directly to the pioneering quantum Monte Carlo studies of Hohenadler \emph{et al.}~\cite{hohenadler2007}, who demonstrated that the relative sign and magnitude of electron and hole phonon couplings dictate the localization of the lattice exciton-polaron. Our Control A calculation explicitly verifies this in the exact diagonalization framework: reversing the relative sign of the couplings (rendering the phonon source even under electron-hole exchange) instantly breaks the exchange selection rule and drives the system into a heavily dressed state, even at perfect mass symmetry. This confirms that the "protected" regime is uniquely tied to the odd-parity (neutral) source combined with the symmetric ground state.

Furthermore, by decoupling the mass ratio from the small-polaron atomic limit via the $\delta$ parameterization, we have isolated the pure symmetry-breaking effect of mass asymmetry. The smooth crossover from a protected exciton to a dressed exciton-polaron as the mass ratio increases provides a lattice-level mechanism that is qualitatively consistent with empirical trends in polar semiconductors. For instance, in mass-symmetric halide perovskites~\cite{miyata2017,johnston2016,naturemat2024}, excitons exhibit sharp zero-phonon lines and resist self-trapping despite strong polar coupling, whereas in highly mass-asymmetric oxides like ZnO and TiO$_2$~\cite{klingshirn2007,song1993}, excitons are heavily dressed and prone to localization. While our 1D Holstein model cannot quantitatively capture the long-range Fr\"ohlich coupling or complex band structures of these real materials, it establishes the fundamental principle that mass asymmetry acts as a tuning parameter for the exchange selection rule, thereby controlling the degree of polaronic dressing.

Perhaps the most striking conceptual result is the paradox of protection. The very exchange symmetry that shields the mass-symmetric exciton from phonon dressing simultaneously denies it the polaronic stabilization energy harvested by free carriers. Consequently, as the electron-phonon coupling increases, the energy of two independent polarons drops rapidly, eventually undercutting the protected exciton. Within the present model, this makes dissociation energetically favorable at sufficiently strong coupling. This mechanism is distinct from phonon-assisted dissociation driven by thermal fluctuations or external fields~\cite{tmd_dissociation}, as it is a purely ground-state energetic competition.

\emph{Limitations and future work.} The present study relies on local (Holstein) coupling and a single Einstein phonon branch. Extending this exact diagonalization framework to include long-range Fr\"ohlich coupling and dispersive phonons will be necessary to capture the full momentum-dependent vertex structure relevant to \emph{ab initio} descriptions of excitonic polarons~\cite{dai2024,ghosh2023,sangalli2019}. Furthermore, computing momentum-resolved spectral functions from these exact wave functions would promote the ground-state proxies used here ($Z_0$, $\langle N_{\rm ph}\rangle$) to fully frequency-resolved observables for direct comparison with ultrafast optical and photoemission experiments.

\section{Conclusion}

We have mapped the non-perturbative regime map of a 1D Holstein-exciton model, separating the physics of internal phonon dressing from the thermodynamics of exciton dissociation. By introducing a parameterization that decouples mass asymmetry from the small-polaron limit, and by performing rigorous control calculations on the phonon source parity, we proved that an exact electron-hole exchange selection rule protects the mass-symmetric, neutral exciton from local polaronic dressing. As mass asymmetry increases, this selection rule is broken, leading to a smooth crossover into a heavily dressed exciton-polaron. 

Conversely, the boundary for phonon-assisted dissociation is governed by a global energy balance between Coulomb binding and polaronic stabilization, and is remarkably insensitive to the internal dressing state. This leads to a fundamental paradox: the symmetry protection that prevents an exciton from dressing simultaneously deprives it of polaronic stabilization, making it thermodynamically vulnerable to dissociation into free polarons at strong coupling. These results provide a unified lattice framework for understanding how constituent mass ratios and interaction strengths cooperatively dictate the photophysical fate of bound electron-hole pairs in polar environments.

\begin{acknowledgments}
The author acknowledges Kenya Education Network (KENET) research services for computing resources.
\end{acknowledgments}

\appendix
\section{Convergence, Controls, and Finite-Size Analysis}\label{app:conv}

\emph{Parameterization and Controls.} The primary results utilize the $\delta$ parameterization ($t_e = t_0 e^\delta, t_h = t_0 e^{-\delta}$) to hold the geometric mean of the hoppings constant, ensuring that variations in dressing are driven by the mass ratio $e^{2\delta}$ rather than the trivial atomic limit $t \to 0$. The control calculations (Control A: even source $g_e=g, g_h=-g$; Control B: $V=0$) were performed at $N=8, n_{\rm tot}=6$ to verify the exchange selection rule and the necessity of the bound state, respectively.

\emph{Finite size.} At fixed truncation $n_{\rm tot}=5$, increasing $N$ from 8 to 10 changes $E_{\rm bind}$ by $\le 0.03$, $Z_0$ by $\le 0.015$, and $\langle N_{\rm ph}\rangle$ by $\le 0.04$ at representative points in the regime map. The signs of $E_{\rm bind}$ and the contrast in $Z_0$ between the protected and dressed regimes are invariant. While a full finite-size scaling to the thermodynamic limit is beyond the scope of this exact diagonalization study, the stability of the level crossing location suggests it is consistent with a dissociation transition in the bulk limit.

\emph{Phonon truncation and reference states.} Protected quantities (Regime I) are converged to $<1\%$ ($Z_0=0.944$, $\langle N_{\rm ph}\rangle=0.10$ at $\delta=0$, $g=2$, $n_{\rm tot}=6$). Dressed states (Regime II and the dissociated regime) carry $O(1)$-$O(4)$ phonons and show residual $n_{\rm tot}$ dependence, as expected for strong local polaronic dressing. The drift of $E_{\rm bind}$ computed with a shared truncation is dominated by the free-polaron references. We therefore converge the exciton sector at $n_{\rm tot}=6$ but the single-polaron references independently at $n_{\rm tot}=10$ (a sector an order of magnitude cheaper), quoting $E_{\rm bind}$ as a difference of independently converged quantities. With this protocol, the dissociation boundary and the energy-balance estimate [Eq.~(3)] are robust.


\begin{figure}[t]
\centering
\includegraphics[width=\columnwidth]{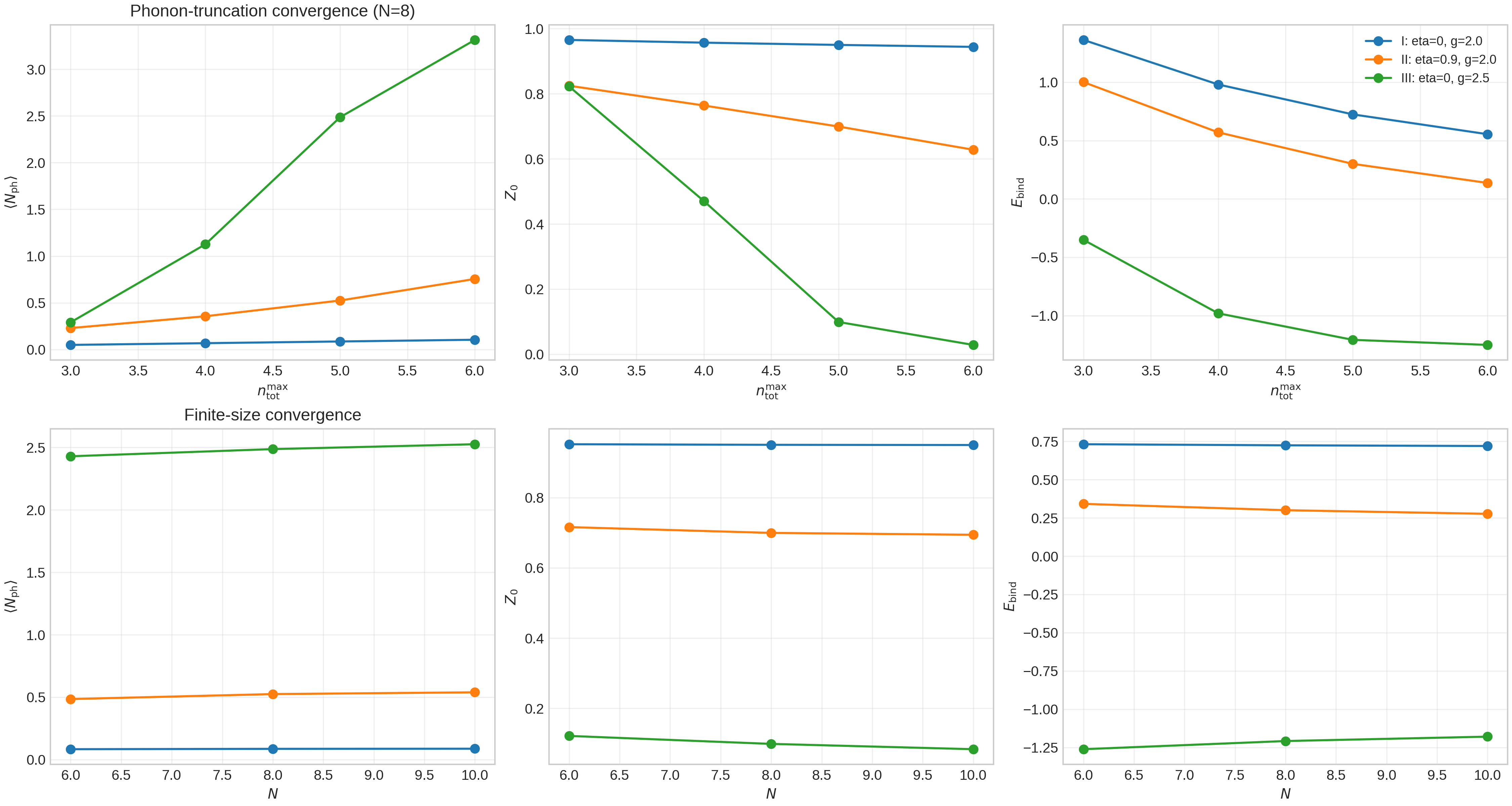}
\caption{Convergence analysis: phonon-truncation sweep at $N=8$ (top) and finite-size sweep at $n_{\rm tot}=5$ (bottom) for the three representative points. The two-tier reference protocol (converging free polarons to $n_{\rm tot}=10$) ensures the binding energies and regime boundaries are robust; see Appendix~\ref{app:conv}.}
\label{fig:convergence}
\end{figure}


\begin{thebibliography}{0}%
\makeatletter
\providecommand \@ifxundefined [1]{%
 \@ifx{#1\undefined}
}%
\providecommand \@ifnum [1]{%
 \ifnum #1\expandafter \@firstoftwo
 \else \expandafter \@secondoftwo
 \fi
}%
\providecommand \@ifx [1]{%
 \ifx #1\expandafter \@firstoftwo
 \else \expandafter \@secondoftwo
 \fi
}%
\providecommand \natexlab [1]{#1}%
\providecommand \enquote  [1]{``#1''}%
\providecommand \bibnamefont  [1]{#1}%
\providecommand \bibfnamefont [1]{#1}%
\providecommand \citenamefont [1]{#1}%
\providecommand \href@noop [0]{\@secondoftwo}%
\providecommand \href [0]{\begingroup \@sanitize@url \@href}%
\providecommand \@href[1]{\@@startlink{#1}\@@href}%
\providecommand \@@href[1]{\endgroup#1\@@endlink}%
\providecommand \@sanitize@url [0]{\catcode `\\12\catcode `\$12\catcode `\&12\catcode `\#12\catcode `\^12\catcode `\_12\catcode `\%12\relax}%
\providecommand \@@startlink[1]{}%
\providecommand \@@endlink[0]{}%
\providecommand \url  [0]{\begingroup\@sanitize@url \@url }%
\providecommand \@url [1]{\endgroup\@href {#1}{\urlprefix }}%
\providecommand \urlprefix  [0]{URL }%
\providecommand \Eprint [0]{\href }%
\providecommand \doibase [0]{https://doi.org/}%
\providecommand \selectlanguage [0]{\@gobble}%
\providecommand \bibinfo  [0]{\@secondoftwo}%
\providecommand \bibfield  [0]{\@secondoftwo}%
\providecommand \translation [1]{[#1]}%
\providecommand \BibitemOpen [0]{}%
\providecommand \bibitemStop [0]{}%
\providecommand \bibitemNoStop [0]{.\EOS\space}%
\providecommand \EOS [0]{\spacefactor3000\relax}%
\providecommand \BibitemShut  [1]{\csname bibitem#1\endcsname}%
\let\auto@bib@innerbib\@empty
\end{thebibliography}%


\begin{thebibliography}{99}

\bibitem{landau1933} L. D. Landau, Phys. Z. Sowjetunion \textbf{3}, 664 (1933).
\bibitem{frohlich1954} H. Fr\"ohlich, Adv. Phys. \textbf{3}, 325 (1954).
\bibitem{holstein1959} T. Holstein, Ann. Phys. (N.Y.) \textbf{8}, 325 (1959); \textbf{8}, 343 (1959).
\bibitem{toyozawa1958} Y. Toyozawa, Prog. Theor. Phys. \textbf{20}, 53 (1958).
\bibitem{rashba1982} E. I. Rashba, in \emph{Modern Problems in Condensed Matter Sciences}, Vol.~2, edited by E. I. Rashba and M. D. Sturge (North-Holland, Amsterdam, 1982), p.~543.
\bibitem{emin2013} D. Emin, \emph{Polarons} (Cambridge University Press, Cambridge, 2013).
\bibitem{alexandrov1995} A. S. Alexandrov and N. F. Mott, \emph{Polarons and Bipolarons} (World Scientific, Singapore, 1995).
\bibitem{song1993} K. S. Song and R. T. Williams, \emph{Self-Trapping of Excitons} (Springer-Verlag, Berlin, 1993).
\bibitem{klingshirn2007} C. Klingshirn, Phys. Status Solidi B \textbf{244}, 3027 (2007).
\bibitem{miyata2017} K. Miyata \emph{et al.}, Sci. Adv. \textbf{3}, e1701217 (2017).
\bibitem{johnston2016} M. B. Johnston and L. M. Herz, Acc. Chem. Res. \textbf{49}, 146 (2016).
\bibitem{sangalli2019} D. Sangalli \emph{et al.}, J. Phys.: Condens. Matter \textbf{31}, 325902 (2019).

\bibitem{hohenadler2007} M. Hohenadler, P. B. Littlewood, and H. Fehske, Phys. Rev. B \textbf{76}, 184303 (2007).
\bibitem{burovski2008} P. Burovski, H. Fehske, and A. S. Mishchenko, Phys. Rev. Lett. \textbf{101}, 116402 (2008).
\bibitem{dai2024} Z. Dai \emph{et al.}, Phys. Rev. B \textbf{109}, 045202 (2024).
\bibitem{ghosh2023} S. Ghosh and F. Paesani, Chem. Sci. \textbf{14}, 13002 (2023).
\bibitem{naturemat2024} X. Li \emph{et al.}, Nat. Mater. \textbf{23}, 1055 (2024).
\bibitem{tmd_dissociation} Y. Zhao \emph{et al.}, Nanoscale \textbf{13}, 5155 (2021).

\bibitem{B} M. O. Atambo,``Symmetry-protected intraband exciton-phonon scattering and its breakdown by mass asymmetry'' arXiv:2609.29634.

\bibitem{C} M. O. Atambo, ``Parity-controlled electron-hole interference in exciton-phonon coupling '' arXiv:2608.25482 .



\end{thebibliography}
\end{document}